\documentclass[pra,twocolumn,aps,showpacs,letterpaper]{revtex4-1}
\usepackage{graphicx}
\usepackage{subfigure}

\usepackage{epsfig}
\usepackage[english]{babel}
\usepackage{color}
\usepackage{tikz}
\usetikzlibrary{shapes,arrows}
\usepackage{amsmath}

\begin{document}

\title{Minimal Effective Gibbs Ansatz (MEGA): A simple protocol for extracting an accurate thermal representation for quantum simulation}

\author{J.~Cohn}
\affiliation{Department of Physics, Georgetown University, Washington, DC 20057, USA}
\affiliation{IBM Almaden Research, San Jose, CA 95120, USA}
\author{F.~Yang}
\affiliation{Departments of Computer Science and Physics, University of California, Berkeley, CA}
\author{K.~Najafi}
\affiliation{Department of Physics, Georgetown University, Washington, DC 20057, USA}
\affiliation{Department of Phyiscs, Virginia Tech. }
\author{B.~Jones}
\affiliation{IBM Almaden Research, San Jose, CA 95120, USA}
\author{J.~K.~Freericks}
\affiliation{Department of Physics, Georgetown University, Washington, DC 20057, USA}

\begin{abstract}
Quantum Gibbs state sampling algorithms generally suffer from either scaling exponentially with system size or requiring specific knowledge of spectral properties \textit{a priori}. Also, these algorithms require a large overhead of bath or scratch/ancilla qubits. We propose a method, termed the minimal effective Gibbs ansatz (MEGA), which uses a quantum computer to determine a minimal ensemble of pure states that accurately reproduce thermal averages of typical observables. This technique employs properties of correlation functions that can be split into a lesser and greater part; here, we primarily focus on single-particle Green's functions. When properly measured, these correlation functions provide a simple test to indicate how close a given pure state or ensemble of pure states are to providing accurate thermal expectation values. Further, we show that when properties such as the eigenstate thermalization hypothesis hold, this approach leads to accurate results with a sparse ensemble of pure states; sometimes only one suffices. We illustrate the ansatz using exact diagonalization simulations on small clusters for the Fermi-Hubbard and Hubbard-like models. Even if MEGA becomes as computationally complex as other Gibbs state samplers, it still gains an advantage due to its ease of implementation without any \textit{a priori} information about the Hamiltonian and in the efficient allocation of available qubits by eliminating bath qubits and using a minimal number of ancilla. 
\end{abstract}
\date{\today}
\maketitle

\section{Introduction} 
In the mid 1990's it was shown that the time evolution of many-body quantum systems can be simulated efficiently on a quantum computer~\cite{abrams1997simulation}. Since then much progress has been made in developing quantum algorithms for simulating these systems~\cite{aharonov2003adiabatic,berry2007efficient,berry2015simulating}. The ability to extract correlation functions, such as single-particle Green's functions which are important for understanding the bulk behavior of condensed-matter systems, have also been developed for quantum computers~\cite{kreula2016few,bauer2016hybrid}. One difficulty, generally overlooked in these algorithms, is that of initial state preparation. While exploring time dynamics will eventually be a straightforward process on an ideal quantum computer, the complexity of preparing physically relevant states can be challenging for certain systems ~\cite{kempe2006complexity}.  

This is especially true when it comes to preparing Gibbs thermal states at low temperature. Certain algorithms are able to achieve quantum Gibbs state preparation, but generally require a large overhead of ancilla or bath qubits and a long run-time~\cite{riera2012thermalization,poulin2009sampling}. Other approaches can be more efficient, but require \textit{a priori} knowledge about specific spectral properties such as correlation lengths or spectral gaps~\cite{yung2012quantum,brandao2016finite,van2017quantum}. Recently, more approximate approaches to Gibbs state sampling have been explored~\cite{endo2018discovering,martyn2018product}

Here we propose a method termed the ``minimal effective Gibbs ansatz" (MEGA), which uses quantum computers to construct a minimal set of pure states that effectively represents a thermal Gibbs state in the sense that it produces accurate thermal expectation values for typical observables. The MEGA works with any correlation function that can be separated into a lesser and greater part. When a system is in thermal equilibrium, these functions can be Fourier transformed from the time domain to the frequency domain. Here, the fluctuation-dissipation theorem schematically gives:
\begin{equation}\label{eq: ratio}
\frac{lesser(\omega)}{greater(\omega)}=-e^{-\beta(\omega-\mu)},
\end{equation} 
for fermionic correlation functions, where $\beta$ is the inverse temperature and $\mu$ is the chemical potential. In this work, we focus on a specific type of correlation function known as a single-particle Green's function.

The MEGA approach requires one to efficiently prepare pure states within a certain energy window, where the ensemble of pure states resemble a mixed state that is diagonal in the energy eigenbasis. Then, using well-known quantum circuits, we extract the lesser and greater parts of the Green's function with respect to each prepared pure state in the ensemble~\cite{kreula2016few,bauer2016hybrid}. Using the known relation of the ratio between the lesser and greater components, given in Eq.~(\ref{eq: ratio}), one can classically extract the optimal $\beta$ and $\mu$ and use this information as an indicator of how well the current ensemble approximates the corresponding Gibbs state.

One advantage of the MEGA lies in its simple implementation and its efficient use of qubits. If one has no prior information as to whether a minimal thermal representation of pure states may exist, one can simply implement the MEGA and test how quickly the results converge. If it does not converge well, then it would be more appropriate to use a different Gibbs state preparation or sampling algorithm. Further, we expect the MEGA to efficiently give a minimal representation in systems where the eigenstate thermalization hypothesis holds, or at temperatures where the system has a finite correlation length~\cite{PhysRevE.50.888,d2016quantum,deutsch2018eigenstate,brandao2016finite,kliesch2014locality}.

The paper is structured as follows. In Sec.II, we briefly review single-particle Green's functions. In Sec.III, we discuss heuristic arguments that support the MEGA being an efficient method, and in Sec.IV, we present numerical simulations. Finally, in Sec.V we give our concluding remarks.

 \section{Single-Particle Green's Functions}
 Single-particle Green's functions are the workhorse of many-body physics. They can be employed to determine a number of properties directly, such as the total energy, double occupancy, kinetic energy, electron filling, etc. In addition, they are required in formulating more complicated response functions like an optical conductivity or a magnetic susceptibility. Here, we will also primarily focus on the lesser and greater Green's functions, which can be seen as a decomposition of the retarded Green's function in the following manner:
 \begin{equation}
G^{>}_{ij\sigma}(t)=-i\langle \hat{c}_{i,\sigma}(t)\hat{c}^{\dagger}_{j,\sigma}(0)\rangle
\end{equation}
\begin{equation}
G^{<}_{ij\sigma}(t)=i\langle \hat{c}^{\dagger}_{j,\sigma}(0)\hat{c}_{i,\sigma}(t)\rangle
\end{equation}
\begin{equation}
G^{R}_{ij\sigma}(t)=\Theta(t)[G^{>}_{ij\sigma}(t)-G^{<}_{ij\sigma}(t)]
\end{equation}
Here the angled brackets represent thermal averaging with respect to the Gibbs state:
\begin{equation}
\rho_G(\beta)=\frac{1}{\mathcal{Z}(\beta)}e^{-\beta\hat{H}}
\end{equation}
where $\mathcal{Z}(\beta)$ is the partition function:
\begin{equation}
\mathcal{Z}(\beta)=\rm{Tr}\{e^{-\beta\hat{H}}\}
\end{equation}
The time dependence of the operators is in the Heisenberg representation. The $\hat{c}_{i,\sigma}(\hat{c}^{\dagger}_{i,\sigma})$ operators represent the Fermionic annihilation (creation ) operators at the $i-th$ site on a lattice for a given $z$-component of the spin, $\sigma\in \{\uparrow,\downarrow\}$. 

When periodic boundary conditions are imposed on the real-space lattice, the creation and annihilation operators can also be represented in momentum space as:
\begin{equation}
\hat{c}_{\vec{k},\sigma}=\frac{1}{\sqrt{L}}\sum_{j=1}^{L}\hat{c}_{j,\sigma}e^{-i\vec{k}\cdot\vec{R}_j}
\end{equation}
where $\vec{k}$ is a reciprocal lattice vector, $\vec{R}_j$ is the real space position vector of the $j-th$ site, and $L$ is the number of lattice sites. 

One can also express the Green's functions in what is known as the Lehmann representation by expanding the trace as a sum over the energy eigenstates (which satisfy $\hat{H}|E_n\rangle=E_n|E_n\rangle$) and inserting a resolution of the identity operator in between the creation and annihilation operators. This is shown below for the lesser Green's function.
\begin{equation}
G^{<}_{ij,\sigma}(t)=\sum_n\frac{e^{-\beta E_n}}{\mathcal{Z}(\beta)}\langle E_n| \hat{c}_{i,\sigma}^{\dagger}e^{i\hat{H}t}\hat{c}_{j,\sigma}e^{-i\hat{H}t}|E_n\rangle
\end{equation}
\begin{equation}\label{eq: Lehmann}
=\sum_n \frac{e^{-\beta E_n}}{\mathcal{Z}(\beta)}\sum_m e^{-i(E_n-E_m)t}|\langle E_n|\hat{c}^{\dagger}_{i,\sigma}|E_m\rangle|^2.
\end{equation}
Fourier transforming Eq.~(\ref{eq: Lehmann}) from the time domain to the frequency domain we likewise obtain:
\begin{equation}
G^{<}_{ij,\sigma}(\omega)=2\pi i\sum_n \frac{e^{-\beta E_n}}{\mathcal{Z}(\beta)}\sum_m \delta(\omega-E_n+E_m)|\langle E_n|\hat{c}^{\dagger}_{i,\sigma}|E_m\rangle|^2
\end{equation}

\begin{figure}[h]
 	\centerline{
		\includegraphics[scale=0.4]{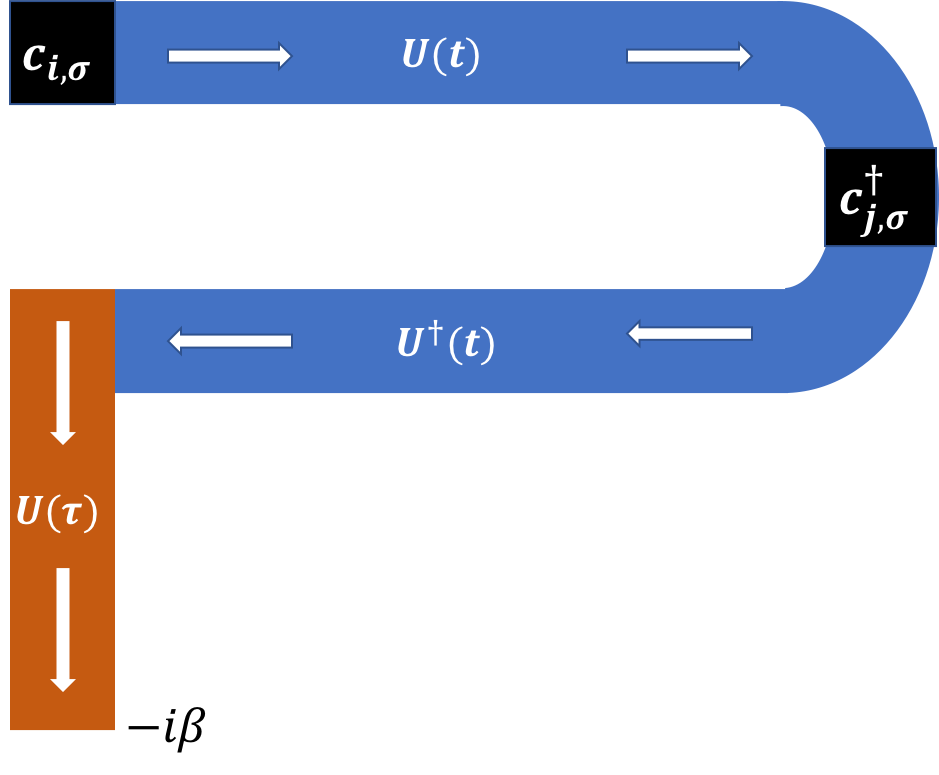}
		}
		\caption{The Keldysh contour for lesser Green's function, with $t>0$. Here one first annihilates a particle at site $i$, then evolves the system for a time $t$, creates a particle at site $j$, evolves backwards in time again for a time $t$, and finally evolves down the imaginary axis to the desired inverse temperature $\beta$. } \label{fig:	Keldysh}
		\end{figure}
An important physical property  that we will also focus on  is the local density of states (per spin) given by
\begin{equation}
A_\sigma(\omega)=-\frac{1}{\pi}{\rm Im}[G^R_{ii,\sigma}(\omega)]=\frac{1}{2\pi}{\rm Im}[G^<_{ii,\sigma}(\omega)].
\end{equation}

The fluctuation-dissipation theorem for Green's functions gives:
\begin{equation}\label{eq: G_ratio}
\frac{G^<_{ij,\sigma}(\omega)}{G^>_{ij,\sigma}(\omega)}=-e^{-\beta(\omega-\mu)}
\end{equation}
which can be easily derived from the grand-canonical ensemble. We can think of the Hamiltonian being shifted by: $\hat{H}\rightarrow\hat{H}-\mu\hat{N}$, with $\hat{N}$ being the total particle number operator, when we work in the grand-canonical ensemble.

\section{Analysis and Heuristics}
Here, we use heuristic arguments to analyze situations in which MEGA is well suited, and examine its limitations. We do not give any rigorous bounds for particular Hamiltonians but rather justify the use of this approach by using physical arguments. MEGA benefits from not needing all the resources required to prepare full Gibbs states when calculating single-particle Green's functions of moderately sized system. We assume the system we describe corresponds to a periodic lattice that is translationally invariant, so that every site is identical. 

In terms of notations, we will use $N$ to determine the number of particles on the lattice. In our analysis below, we will be discussing different partitions of the lattice and of their corresponding Hilbert spaces to analyze the ansatz. Specific partitions of the lattice will be represented by capitol letters such as $A,B,C...$ and their associated Hilbert spaces as $\mathcal{H}^A,\mathcal{H}^B,\mathcal{H}^C...$, with the exception of $D$ which will be used to represent the spatial dimension of the lattice. We will also use $G_{ij,\sigma}(t)$, without a specific superscript, as a general reference that applies to all Green's functions defined above.

The aim of MEGA is to find a minimal set of pure states, $\{|\psi_i\rangle\}$, in which ensemble expectation values of typical thermodynamic observables yield accurate approximations to those determined from the corresponding Gibbs state. The minimal ensemble needs to be stationary to give proper thermal results. We can expand each state as, $|\psi_i\rangle=\sum_n\alpha^i_n|E_n\rangle$, where $\alpha^i_n=\langle E_n|\psi_i\rangle$. Here,  if we take a uniform ensemble average of these states we end up with:
\begin{multline}
\frac{1}{\mathcal{N}}\sum_i|\psi_i\rangle\langle\psi_i|\\=\frac{1}{\mathcal{N}}\sum_i\Bigg[\sum_n|\alpha^i_n|^2|E_n\rangle\langle E_n|+\sum_{n,m\neq n}\alpha^i_n(\alpha^i_m)^*|E_n\rangle\langle E_m|\Bigg]
\end{multline}
where $\mathcal{N}$ is the number of states in the ensemble. One can also weight the states in the MEGA ensemble nonuniformly, as we do when we introduce Boltzmann weights, instead of using a uniform distribution. 

In order for the ensemble to be stationary, since statistical ensembles are time-independent, the off-diagonal elements must average to zero. To this effect, the set of states defined in the MEGA can either be a distribution over a small set of energy eigenstates or they can be an arbitrary set of pure states, whose ensemble average sum produces a distribution over a set of eigenstates within some energy window. One can also loosen the restrictions on stationary to ``approximately stationary" by bounding the quantum fluctuations of the target observable given by the MEGA ensemble. There are two simple ways to make the ensemble stationary: (i) one can average over time so that the off-diagonal elements are smaller than some given tolerance, or (ii) one can project the density matrix onto its diagonal elements in the energy basis. While the former method always is possible, it can entail a large number of measurements if the time domain being covered is large. Projective measurements in the energy basis will require using some form of phase estimation, and may be problematic for low-depth circuits. We anticipate that in the short term either one or both of these options will become viable on available quantum hardware.

 One drawback of MEGA is that we cannot dial in specific temperatures, instead one generally has approximate bounds in terms of what consists of low vs. high energy for a given Hamiltonian. By preparing a state in a narrow energy window, one can then extract the effective temperature via our post-processing procedure employing the ratio of the lesser and greater Green's functions in frequency space. 
 
 There are also many ways to realize the ensemble of pure states. One could use a state preparation procedure based off repeatedly preparing the same state, $|\psi_0\rangle$, which has an overlap with many eigenstates within a given energy window. After each round of state preparation, projective energy measurements are executed, so that the state collapses to one of the energy eigenstates in the superposition. Here, post-selection can also be utilized to discard any eigenstates with energy outside of the desired window. In this case, one would have to examine the cost of performing projective energy measurements. If the populated energy states are not too dense, then one may be able to use iterative phase estimation with a single ancilla qubit to collapse to a single eigenstate~\cite{dobvsivcek2007arbitrary,o2018quantum}. When the states are more densely populated, then more aniclla qubits would have to be utilized in order for the process to be efficient~\cite{kitaev1995quantum,abrams1999quantum}.   
 
 Instead of using projective energy measurements one could use an ensemble based off evolving a single   pure state for different time periods, $\Big\{|\psi_i\rangle, \hat{U}(\Delta t)|\psi_i\rangle, \hat{U}(2\Delta t)|\psi_i\rangle,...\Big\}$. This time averaged ensemble will be equivalent to the projective measurements as long as one chooses the proper time window. Finally, one could attempt to randomly prepare states within a given energy window, extracting the Green's functions for each random pure state. Choosing an appropriate sample of states would force the ensemble to be stationary or approximately stationary. 
 
The precise methodology needed to create states within a given energy window will depend on the specific system as well as the hardware and resource limitations. We do not spell out a particular algorithm here, but one can choose from a variety of known methods such as adiabatic state preparation, the variational quantum eigensolver (VQE)~\cite{peruzzo2014variational,mcclean2016theory}, the quantum approximate optimization algorithm (QAOA)~\cite{farhi2014quantum}, quantum walk algorithms, or amplitude amplifications to construct the best approach with the given system and resource limitations~\cite{shenvi2003quantum,brassard2002quantum,poulin2017fast}. We also note that isolating a narrow, low-energy window can still be exponentially hard for certain problems, but we expect other Gibbs state preparation algorithms to suffer here as well~\cite{kempe2006complexity}.
 
The advantage of using the MEGA is in the simple implementation and the efficient use of the available qubits. Fermionic systems usually require $2L$ qubits per lattice via the Jordan-Wigner mapping to a corresponding spin Hamiltonian. There also exist parity mappings such as the Bravyi-Kitaev map that only require $L$ qubits~\cite{bravyi2002fermionic}. Also, correlation functions such as single-particle Green's functions can be extracted using a single ancilla qubit~\cite{kreula2016few,bauer2016hybrid}. Given this information the MEGA should require at most $2L+1$ qubits. 

An outline of the MEGA procedure is as follows:
\begin{enumerate}
\item Prepare $|\psi_i\rangle$ with $N$ electrons and within an appropriate energy window. (One has the option of additionally employing projective measurements here depending on available resources to remove states that fall outside the desired window).
\item Repeatedly use the same state preparation procedure to measure $G^<_{ij,\sigma}(t)$ and $G^>_{ij,\sigma}(t)$ at a series of points in time. Extend the time points far enough out that the Green's function can have its tail fit to an exponential or power-law decay. (Negative times can be extracted by using the relation that the imaginary parts of the lesser and greater Green's functions are symmetric about $t=0$ and the real parts are anti-symmetric about $t=0$.)
\item On a classical computer, perform a Fourier transform from time to frequency, approximating the real time Green's functions by the fit tail for large enough times. Extract a least squares fit of $\beta$ and $\mu$ from the Eq.~(\ref{eq: G_ratio}).
\item If the least squares fit lies below a given threshold, terminate the calculation. Otherwise, return to step 1, and prepare $|\psi_{i+1}\rangle$, possibly using least squares fits of the current set of states to inform the state preparation procedure for $|\psi_{i+1}\rangle$.
\end{enumerate}
Note that there is no guarantee that MEGA will produce a sparse representation of a thermal Gibbs state, but the advantage here is that one can implement the MEGA protocol without any prior knowledge and observe how quickly the fit converges. 

The ratio of the lesser to greater Green's function, which we employ to test the accuracy of the MEGA for a given calculation, is derived in the grand-canonical ensemble. But the calculation procedure described above worked with a fixed filling of the electrons. This is fine for a large enough system, because the microcanonical, canonical, and grand-canonical ensembles all yield the same results~\cite{brandao2015equivalence,tasaki2018local}. But for finite sized lattices, one may do better by adding states with different fermion fillings, or by weighting states in the ensemble by Boltzmann factors to improve the convergence of the MEGA. One additional limitation is of numerical precision in instances where there is a gap in the local density of states. Here, we run into the problem of trying to divide two numbers that are approximately zero in the gap region.  

Since the ratio of the lesser to the greater Green's function is employed in a post-processing step, we can modify the procedure if we run into numerical instabilities. For example, we can find a similar relation by using density-density correlations such as:
\begin{equation}\label{eq: corr_R}
\mathcal{C}^R_{ij,\sigma}(t)=-i\Theta(t)\langle[\hat{n}_{i,\sigma}(t),\hat{n}_{j,\sigma}(0)]\rangle
\end{equation}
where we can define:
\begin{equation}
\mathcal{C}^<_{ij,\sigma}(t)=-i\langle\hat{n}_{i,\sigma}(t)\hat{n}_{j,\sigma}(0)\rangle 
\end{equation}
and
\begin{equation}
\mathcal{C}^>_{ij,\sigma}(t)=-i\langle\hat{n}_{j,\sigma}(0)\hat{n}_{i,\sigma}(t)\rangle
\end{equation}
Fourier transforming these correlators to the frequency domain yields:
\begin{equation}\label{eq: corr_ratio}
\frac{\mathcal{C}^<_{ij,\sigma}(\omega)}{\mathcal{C}^>_{ij,\sigma}(\omega)}=e^{-\beta\omega}.
\end{equation}
These correlators allow for a direct canonical ensemble fit because the operators, $\hat{n}_{i,\sigma}$, preserve the filling of a given eigenstate. The local lesser and greater correlators here have the same cost to implement as the local lesser and greater Green's functions on a quantum computer. More generally, if one wants the MEGA to approximate thermal properties on larger regions, then one should construct proper correlation function whose underlying operators are supported across that specific domain.

To minimize the gate depth for near term use a variant of the VQE algorithm (or, when achievable, preparing simple product states) would most likely be the best choice to prepare a given ensemble of the $|\psi_i\rangle$'s.  The main limiting factor for near term use will then come from the required time evolution to extract the Green's functions. Here, we will need to extract the Green's functions for many time steps extending out to at least a characteristic decay time $t=t_d$ (where the Green's function becomes vanishingly small). Optimistically, we expect the circuit depth here to scale linearly with the number of Trotter steps and hence linearly with $t_d$. The depth of each Trotter step will scale polynomially on the number of sites/orbitals($\mathcal{O}(L^{1/2})-\mathcal{O}(L^4)$) whose exact exponent again depends on the specific system being simulated~\cite{jiang2018quantum,motta2018low}. With these circuit depth requirements, we expect the MEGA to be applicable once circuit depths required for modest time evolution can be reached~\cite{haah2018quantum}. The MEGA approach is also limited by the complexity of preparing states within a narrow energy window, which can be difficult for certain systems, but this complexity will also limit other Gibbs state preparation algorithms as well.

Now, obviously we want the number of states, in the extracted ensemble, to scale much less than the number of relevant energy eigenstates in the corresponding Gibbs state. Otherwise, this approach would be equivalent to doing full diagonalization, which would have a large exponential scaling. A naive reduction of the number of relevant states is through ensemble equivalence. Within the grand-canonical ensemble picture the fluctuations, $\delta E$ and $\delta N$, scale as $1/\sqrt{N}$. The variances, $\delta E$ and $\delta N$, will also have a dependence on temperature which becomes relevant with small system sizes. So, if we are working with a fixed filling, $n=N/L$, then the microcanonical, canonical, and grand-canonical ensembles should become equivalent as the systems size becomes large. 

In this limit, one can sample from a narrow energy window with fixed particle number.  While the microcanonical ensemble offers a slight reduction in the number of relevant states compared to the other statistical ensembles, it will still contain an exponential number in the thermodynamic limit. We also note that below certain temperatures ensemble equivalence may not be well defined for certain systems if the relevant energy eigenstates have zero energy density, $(E_n-E_0)/N$ in the thermodynamic limit. It is possible, with certain systems in this temperature range, that thermodynamic quantities remain dependent on the ensemble, even in the thermodynamic limit. 

The ideal setting, where MEGA would yield a sparse ensemble, is in systems where the eigenstate thermalization hypothesis (ETH) holds. The ETH ansatz states that when a random pure state is chosen from a superposition of states originating from a narrow energy window (lying away from the edges of the spectrum of generic non-integrable systems), then the matrix elements of typical few-body observables take the form~\cite{PhysRevE.50.888}:
\begin{equation}
\langle E_m|\hat{O}|E_n\rangle=O_{mc}(\bar{E})\delta_{mn}+e^{-\frac{S(\bar{E})}{2}}f_O(\bar{E},\omega)R_{mn}.
\end{equation}
Here, $O_{mc}(\bar{E})$ is the microcanonical average of the observable $\hat{O}$ centered at the average energy $\bar{E}$ of the narrow energy window and $S(\bar{E})$ is the thermodynamic entropy defined by $\exp[S(\bar{E})]=\bar{E}\sum_n^{\prime}\delta_{\epsilon}(\bar{E}-E_n)$, where the restricted sum is over the number of states within a smeared delta function centered at $\bar{E}$. $f_O(\bar{E},\omega)$ is a smooth function of its arguments with $\omega=E_n-E_m$ and $R_{mn}$ is a random number with zero mean and unit variance. Here one can see that as the number of states within the energy window become exponentially large then the fluctuations about the microcanonical ensemble become exponentially suppressed. Note that there is no requirement that the typical few-body observables need to be local in space, as momentum occupation is generally used in numerical studies of ETH~\cite{PhysRevA.43.2046,rigol2008m,rigol2008thermalization}. 

When ETH holds, a single energy eigenstate from the bulk of the spectrum is expected to give the same expectation values for local observables as a corresponding Gibbs state with the same average energy i.e.
\begin{equation}
\langle E_n|\hat{O}_{loc}|E_n\rangle\simeq\rm{Tr}(\rho_G(\beta)\hat{O}_{loc})
\end{equation}  
where $\beta$ is chosen so that the equality, $\langle E_n|\hat{H}|E_n\rangle=Tr(\rho_G(\beta)\hat{H})$, holds. The ETH can also be interpreted as having a single eigenstate from the bulk act as a bath for local subspaces~\cite{kim2014testing}. Here, even though the single eigenstate and corresponding Gibbs state are globally very different, when one examines their reduced density matrices on local subspaces the representations become equivalent in the thermodynamic limit. While it is not known for which operators the ETH holds, some arguments based on conformal field theory allow for the observables to be supported on a subsystem as large as half the size of the full system~\cite{garrison2018does,cardy2014thermalization}.  When the ETH holds for all (strong-ETH) or typical (weak-ETH) energy eigenstates within the desired energy window, this should allow for a sparse ensemble size for the MEGA. The downside here is that ETH is not expected to hold for the edges of the spectrum as the energy density does not remain finite in the thermodynamic limit, so there would be a lower bound on the temperature that can be reached by a MEGA-based simulation. One potential application of the MEGA is the ability to locate where the energy spectrum transitions from zero-energy density to finite-energy density in large systems. 

Another example where we expect a sparse ensemble is in systems where the correlations decay exponentially. In this case, we can associate a temperature-dependent correlation length, $\xi(\beta)$ to the Gibbs state of a specific system. When $\xi(\beta)$ has a finite length, then sampling from a Gibbs state prepared on a system equal to the size of the observable of interest plus the correlation length efficiently yields the behavior of a Gibbs state prepared on the same system in the thermodynamic limit~\cite{kliesch2014locality,brandao2016finite}. 

As shown schematically in~Fig.~\ref{fig: corr_len}, we decompose the Hilbert space of our closed system as, $\mathcal{H}^S=\mathcal{H}^A\otimes\mathcal{H}^B\otimes\mathcal{H}^C$, where $\mathcal{H}^A$ is the Hilbert space supported by the local operator, $\mathcal{H}^B$ is the Hilbert space described by a ring of width $\xi(\beta)$ around region-$A$, and $\mathcal{H}^C$ is the Hilbert space of the rest of the system. Preparing a Gibbs state with inverse temperature $\beta$ on the space $\mathcal{H}^A\otimes\mathcal{H}^B$ gives the same results for observables supported solely in region $A$,  as one would get if the full system Gibbs state was prepared. Here, the temperature-dependent correlation length should, in principle, place bounds on the minimum size of the ensemble achievable from the MEGA approach. 

\begin{figure}[h!]
 	\centerline{
		\includegraphics[scale=0.4]{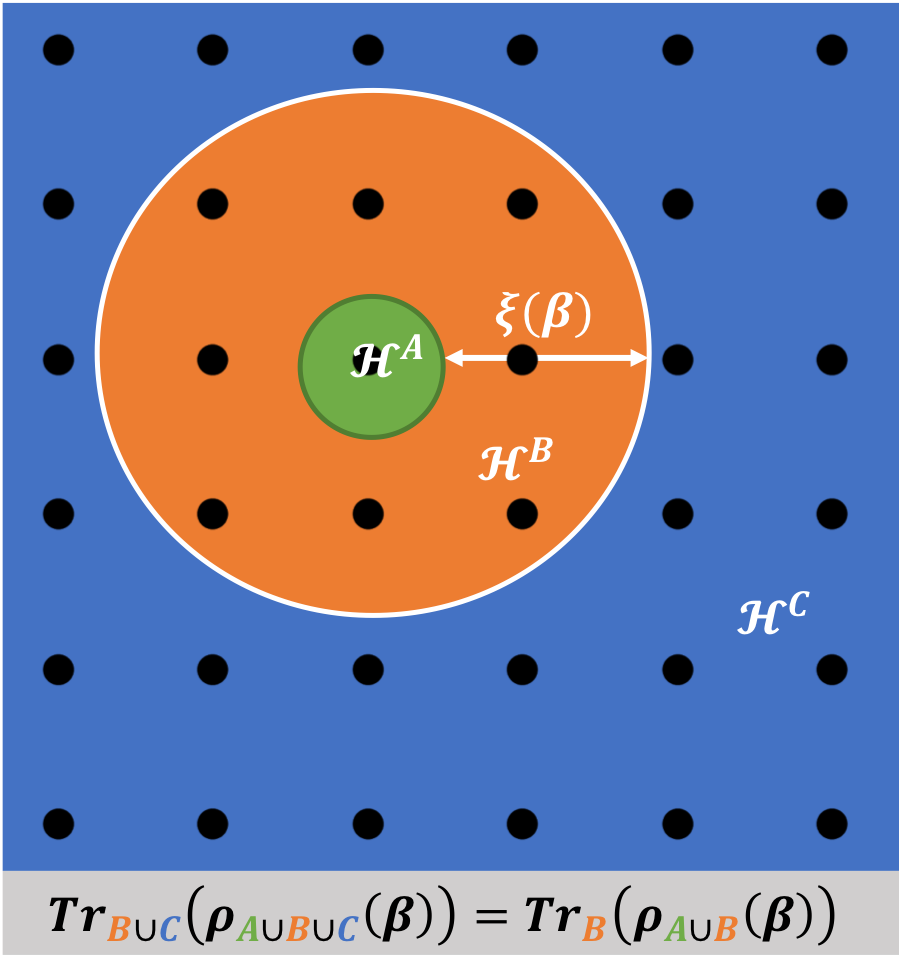}
		}
		\caption{Diagram of lattice decomposed into three disjoint regions: $A$ (green), $B$ (orange), and $C$ (blue). Region $A$, represents the Hilbert space which support the local operator. Region $B$, represents a shielding region around $A$ whose width is the size of the correlation length, $\xi(\beta)$, and Region $C$ represents the rest of the system. Here a Gibbs state prepared on the region $A\cup B$ with inverse temperature, $\beta$, is locally equivalent on $\mathcal{H}^A$, to a Gibbs state prepared on the region $A\cup B\cup C$ with the same inverse temperature.} \label{fig:	corr_len}
		\end{figure}

\begin{figure}[h!]
 	\centerline{
		\includegraphics[scale=0.27]{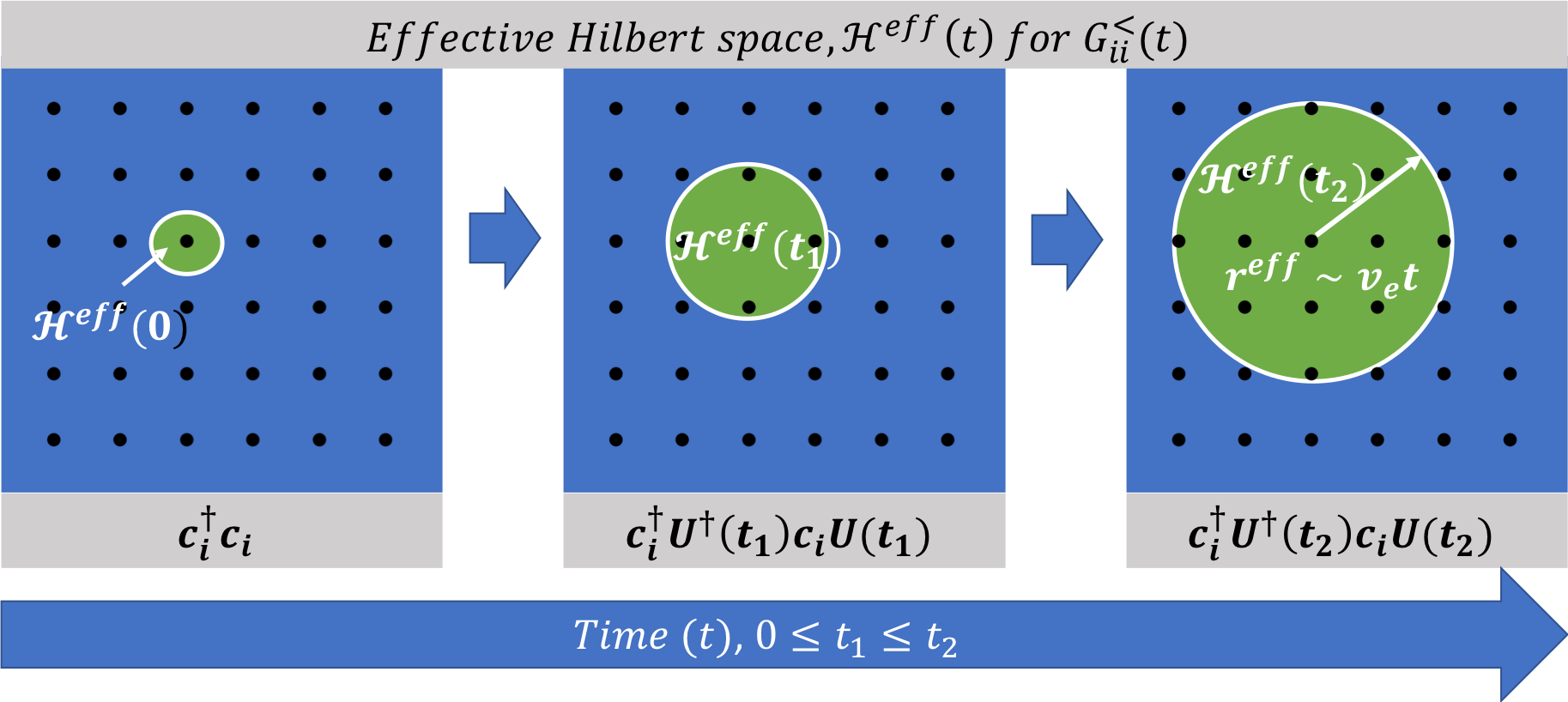}
		}
		\caption{Diagram depicting the growth of the effective Hilbert space of the unequal time operator, $\hat{c}^{\dagger}_{i,\sigma}\hat{U}^{\dagger}(t)\hat{c}_{i,\sigma}\hat{U}(t)$, used in the local lesser Green's function. A conservative estimate is that the effective Hilbert space growth is bounded by the entanglement velocity,$v_e$.} \label{fig:	eff_hilb}
\end{figure}

The local Green's function initially depends just on the electron density on lattice site $i$. As time increases the operators grow radially outward in time. There are three characteristic speeds that govern this growth.The first is the Lieb-Robinson velocity, $v_l$, which determines the casual light cone, bounding how quickly an initial localized perturbation can affect the rest of the system~\cite{lieb1972finite}. The butterfly velocity, $v_b$, describes the speed at which the wavefront of an initially local operator, $\hat{c}_{i,\sigma}(t=0)$, grows radially outward in space as a function of time ($\hat{c}_{i,\sigma}\rightarrow\hat{U}^{\dagger}(t)\hat{c}_{i,\sigma}\hat{U}(t)$) ~\cite{maldacena2016bound,khemani2017operator,von2018operator}. The butterfly velocity is determined from an out-of-time-ordered correlation function originally defined in~\cite{larkin1969quasiclassical}. The third is the entanglement growth, $v_e$, again generated by $U^\dagger(t)\hat{c}_{i,\sigma}U(t)$, which estimates the rate of bipartite entanglement production when this unequal time operator acts on an initial product state (see ~Fig.\ref{fig:		eff_hilb})~\cite{von2018operator}. These characteristic speeds are expected to satisfy: $v_e\leq v_b\leq v_l$ and are system dependent, where local interactions, geometry, and conserved quantities play a large role in determining their form.  Also, for thermalizing systems, these Green's functions have a system dependent decay time, $t_d$ which depends on temperature. Given this information, if one uses local single-particle Green's functions to converge an ensemble of states with MEGA, this ensemble should yield accurate thermal expectation values for operators supported on an effective Hilbert space, qualitatively bounded by lattice sites within a radius $\sim v_{e}t_d$ from site-$i$.

\section{Numerical Results}
To test the validity of this approach, we focus on the repulsive 1-D Fermi-Hubbard model and its variants~\cite{hubbard1963electron}. This well known model aims to minimally account for the electron correlations by imposing an interaction that repels two electrons of opposite spin only when they are on the same site. The Hamiltonian is given by:
\begin{equation}
\hat{H}_{Hubb}=-t\sum_{i,\sigma}(c^{\dagger}_{i,\sigma}c_{i+1,\sigma}+h.c.)+U\sum_in_{i,\uparrow}n_{i,\downarrow}
\end{equation}
where, $t$ is the strength of the electron hopping,  and $U$ is the on-site repulsion term. We note the redundant use of notation here where the parameter, $t$, is used to represent both the energy scale of the hopping term and time. Also the parameter $U$ is similar to the time evolution operator $\hat{U}(t)$, which are differentiated by the operator symbol in the time-evolution operator. The context in the text should clarify the intended interpretation of these symbols.    

In one dimension, the model is integrable, and can be solved by the Bethe ansatz, so it is not expected to thermalize to the proper Gibbs ensemble because of the macroscopic number of symmetries the model exhibits. Nevertheless, we show, adjusting certain parameters of this model allows us to predict the effectiveness of the MEGA protocol in larger nonintegrable systems. For concreteness, we also add integrability breaking terms to the Fermi-Hubbard model and compare the performance when these terms are added. When the new terms are added the Hamiltonian becomes
\begin{equation}
\hat{H}=\hat{H}_{Hubb}+\hat{H}^{\prime}
\end{equation}
where
\begin{equation}
\hat{H}^{\prime}=-t^{\prime}\sum_{i,\sigma}(\hat{c}^{\dagger}_{i,\sigma}\hat{c}_{i+2,\sigma}+h.c.)+U^{\prime}\sum_{i}\hat{N}_{i}\hat{N}_{i+1}
\end{equation}
and $\hat{N}_{i,\sigma}=(\hat{n}_{i,\uparrow}+\hat{n}_{i,\downarrow})$.

\begin{figure}[h!]
 	\centerline{
		\includegraphics[scale=0.4]{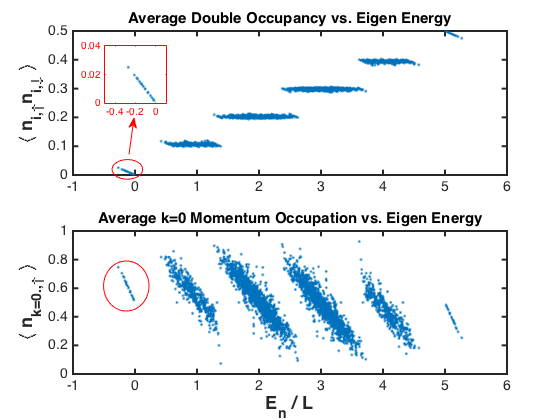}
		}
		\caption{Scatter plots of the average double occupancy $(\langle \hat{n}_{i,\uparrow}\hat{n}_{i,\downarrow}\rangle)$ and $k=0$ momentum occupation $(\langle \hat{n}_{k=0,\sigma}\rangle)$ with respect to each energy eigenstate. One can see that when restricted spin band (outlined in red) these observables are smooth monotonically decreasing functions of eigenstate energy.} \label{fig:	ETH}
		\end{figure}

To begin, we will examine the half-filled 1-D Hubbard model with periodic boundary conditions with a large on-site interaction of $U/t=10,t^{\prime}=U^{\prime}=0$. This specific case is interesting because it exemplifies the ideal behavior of a system obeying the ETH. As one can see in Fig.~\ref{fig:	ETH}, when restricted to the first energy band (spin band), both the double occupancy and the $k=0$-momentum become smooth functions of the eigenstate energy. This is indicative of the strong-ETH in the extreme sense, where every eigenstate is typical. The spectrum as a whole does not obey the ETH, so these results do not indicate physical behavior in the thermodynamic limit. Nevertheless this behavior in the lowest band should give insight into the performance of these approximations in an ideal setting. 

In general, we are more interested in simulating strongly correlated electrons rather than weakly correlated electrons, because weak correlation is amenable to many classical numerical techniques. Choosing low-energy states is relatively simple here, because at infinite interaction, there are no double occupancies at and below half filling. These states are also easy to generate on a quantum computer as product states. So, our strategy is to initialize the system in a state with no double occupancy, ramp the state adiabatically from infinite interaction to finite interaction, and employ such a state as one of the states in the MEGA ensemble.

For these simulations, we employ exact diagonalization and use a MEGA consisting of the two Ne\'{e}l states, each time evolved with a time-dependent Hamiltonian. We initially set the interaction energy to $U/t=500$, making $t$ our energy scale. We also set $\hbar=k_b=1$. We then evolve the system with a time dependent interaction energy that ramps from $U/t=500$ to $U/t=10$ given by the time evolution operator of:
\begin{equation}
\hat{U}_{prep}(t)=\mathcal{T}\Big\{\exp\big[-i\int_{0}^{t}dt^{\prime}\hat{H}_{Hubb}(t^{\prime})\big]\Big\}
\end{equation}
where the time dependence of the interaction energy in the Hubbard Hamiltonian is given by $U(t)=490e^{-t/5}+10$. Here, our final set consists of:
\begin{equation}
 \Big\{\hat{U}_{prep}(\tau)|\uparrow\downarrow\uparrow...\rangle,\hat{U}_{prep}(\tau)|\downarrow\uparrow\downarrow,...\rangle\Big\},
 \end{equation}
  where $\tau$ is the total ramp time.
. We choose to start with the simple Ne\'{e}l state because it is a product state in the localized basis, which would be trivial to prepare on a quantum computer.  The Ne\'{e}l state is one of the ${N}\choose{N/2}$ degenerate product states in the ground state at $U=\infty$. At large but finite $U$,  the Ne\'{e}l states will have an overlap with the ground state and a couple other low-lying energy states. When ramping down to a smaller $U$ we are guaranteed to stay in the spin band as long as we ramp slow enough. One could also ramp up from $U/t=0$, but initializing a quantum computer to the Fermi sea is a more complicated circuit. 
\begin{figure}[h!]
 	\centerline{
		\includegraphics[scale=0.4]{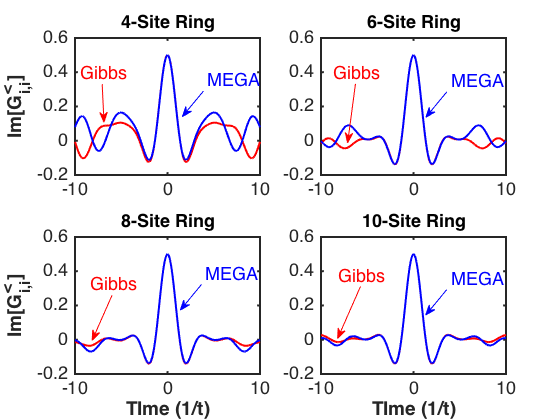}
		}
	\caption{Imaginary parts of the local lesser Green's function calculated with the MEGA approximation (blue curve) and with the exact Gibbs state (red curve). Here we see that the MEGA becomes accurate at later and later times as the system size increases. } \label{fig:	GF_less}
\end{figure}	

 \begin{figure}[h!]
 	\centerline{
		\includegraphics[scale=0.4]{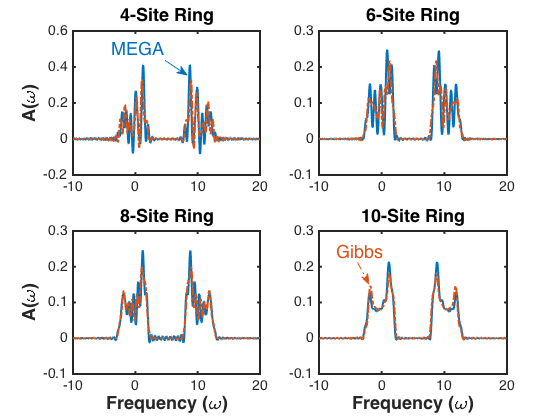}
		}
		\caption{Local density of states for $L=4,6,8,10$. The gap is still identifiable for the 4-site ring but each band quickly converges to the true result as the system size increases.} \label{fig:	Ldos}
 \end{figure}
 
\begin{figure}[h!]
 	\centerline{
		\includegraphics[scale=0.4]{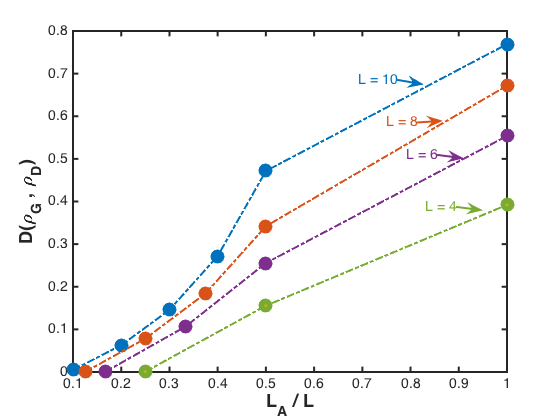}
		}
		\caption{Trace distance between the MEGA ensemble and corresponding Gibbs state's reduced density matrices for $L=4,6,8,10$. From this trend, we see that the larger the system size, the faster these ensembles start converging on subspaces less than half the system size} \label{fig:	trace_dist}
 \end{figure}

 The trends in~Fig. \ref{fig:	GF_less} and~Fig. \ref{fig:	Ldos} show what we would naively expect when examining this ideal system. The local Green's functions, when approximated by just two states, resemble the exact results for longer and longer times as we increase the system size. While the finite size effects prevent the the Green's function from truly decaying to zero, we can interpret here that the effective Hilbert space for $t<t_d$, for a given $t_d$, becomes a smaller and smaller fraction of the total Hilbert space as the size of the system increases. Since we are working in a regime where the ETH holds, the approximations for the times $t<t_d$ become better as the fraction of their support on the total Hilbert space becomes smaller. This is further exemplified in ~Fig. \ref{fig:	trace_dist}. Here we show the trace distance, defined as:
 \begin{equation}
 D(\rho,\rho^{\prime})=\frac{1}{2}tr(|\rho-\rho^{\prime}|),
 \end{equation}
between the Gibbs state and the MEGA. When we examine the trace distance between the reduced density matrix of these states, we can see that they start to become equivalent on small subsystems. In particular, we notice that in the 10-site system, the trace distance between these two states is large when examining the whole system, but when the subsystem is less than half the system size, these two start to rapidly become close to one another. Note that having a small trace distance between the MEGA and the Gibbs state is a sufficient condition for MEGA to be accurate, but is not necessary.
 
\begin{figure}[h!]
 		
 		\centerline{
		\subfigure[]{\includegraphics[scale=0.4]{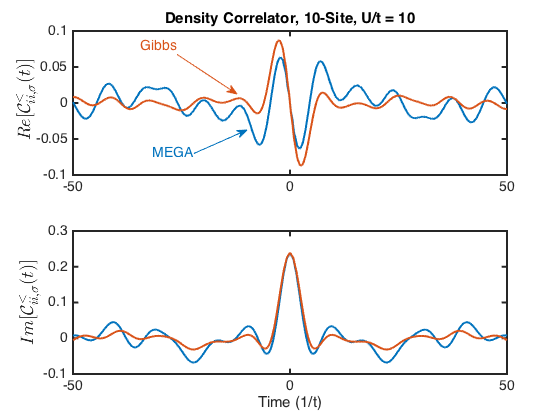}}
		}

	\centerline{
		\subfigure[]{\includegraphics[scale=0.4]{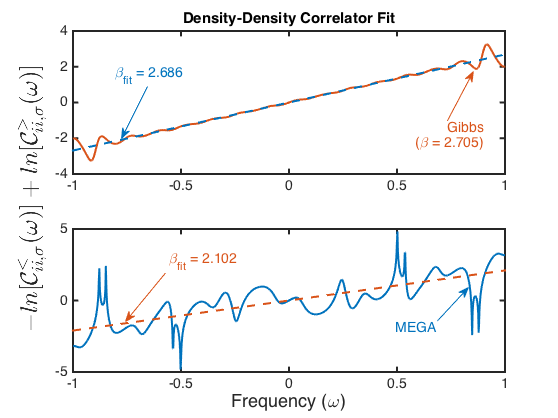}}
		}
		\caption{Canonical Gibbs fit using the density-density correlator instead of the single-particle Green's functions for a 10-site system with $U/t=10$. (a) The MEGA approximation used here, $\Big\{\hat{U}_{prep}(\tau)|\uparrow\downarrow\uparrow...\rangle,\hat{U}_{prep}(\tau)|\downarrow\uparrow\downarrow,...\rangle\Big\}
$, diverges from the corresponding Gibbs state at earlier times for this correlator than with single-particle Green's functions. (b) The figure also shows that the effective temperature can be properly extracted when a large enough ensemble is used. We see that the current MEGA ensemble exhibits large fluctuations and the fit for the effective temperature yields $\beta=2.1$ where the proper Gibbs state ,at the same energy of the MEGA ensemble, has an inverse temperature of $\beta=2.7$. This indicates that the current ensemble used for the MEGA has not properly converged and will need a larger set of states.} \label{fig:	dens_ratio}

		\end{figure}

 Unfortunately, the large gap in the local density of states makes extracting the effective temperature numerical unstable. To get around this problem, we use the density-density correlator defined in Eqs.~(\ref{eq: corr_R}-\ref{eq: corr_ratio}). As shown in Fig.~\ref{fig:	dens_ratio} while this correlator is capable of extracting the correct temperature, it requires a larger set of states in our MEGA approach for the results to properly converge. When using our set of two states for the MEGA approach the results diverge more quickly in the real time domain, which leads a noisy fit when extracting the temperature. Fig.~\ref{fig:	dens_ratio} shows that within this MEGA ensemble there is still a large error present in the temperature extraction. This earlier divergence is most likely due to faster operator spreading of $\hat{U}^{\dagger}(t)\hat{n}_{i,\sigma}\hat{U}(t)$ used in the density-density correlator than in the operator, $\hat{U}^{\dagger}(t)\hat{c}_{i,\sigma}\hat{U}(t)$ used for the single-particle Green's functions. 
 
 This scenario demonstrates the various trade-offs between the different functions that can be used to verify an accurate MEGA ensemble, and how we extract the effective temperature. Here, we see that a small MEGA ensemble accurately converges for single-particle Green's functions, but the gap in the local density of states prevents a proper test of convergence due to numerical precision errors. On the other hand, the density-density correlator is capable of accurately testing for convergence but requires the MEGA to use a larger set of states. Ideally, this small set of states in the MEGA ensemble should become more accurate when using the density-density correlator as we increase the system size as predicted by the ETH ansatz.

\begin{figure}[h!]
 		\centerline{
		\subfigure[]{\includegraphics[scale=0.4]{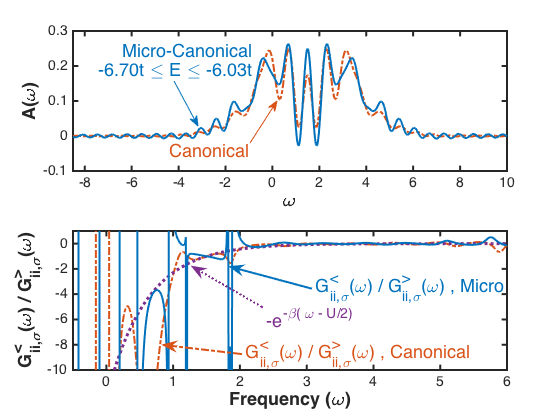}}
		}
	\centerline{
		\subfigure[]{\includegraphics[scale=0.4]{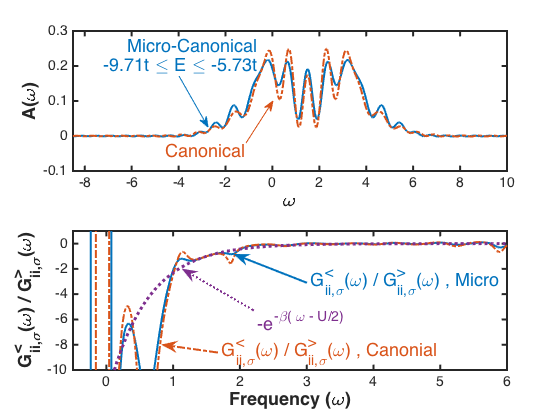}}
		}
		\caption{Local density of states and ratio of local lesser and greater Green's functions for a 10-site system with $U/t=3$. The systems are probed at a temperature of $T=0.6$ and we examine the results of the canonical Gibbs state with two different microcanonical windows. (a) The smaller window has noisier results making it difficult to extract the effective temperature. (b) The fit becomes easier when moving to the larger energy window. } \label{fig:	GF_ratio}
		\end{figure}

 To demonstrate tests for convergence and temperature extraction with single-particle Green's functions, we work with a half-filled 10-site 1-D Hubbard model with $U/t=3$, where there is no longer a gap in the local density of states. Here the spin band is no longer separated from the rest of the spectrum, so ETH effects no longer hold, eliminating the ability of sparse window sampling to efficiently describe the thermal behavior. Figure \ref{fig:	GF_ratio} compares the ratio of $G^<_{ii,\sigma}(\omega)/G^>_{ii,\sigma}(\omega)$ for the Gibbs state at $T=0.6$ to two corresponding microcanonical windows. The first is a small window with energy ranging from $-6.70t\leq E\leq-6.03t$ and the second is a larger window with a range: $-9.71t\leq E\leq-5.73t$. As we can see the small window does not give a good fit to the correct temperature and chemical potential, but the larger window clearly has a more stable fit. The Gibbs state here is represented by the canonical ensemble, so we can also see in Fig.~\ref{fig:	GF_ratio} that the canonical ensemble has not quite converged with the grand-canonical. The rapid oscillations are caused by the numerical instability when both the lesser and greater Green's functions approach zero. 
 
These results show an example of where the MEGA is able to eventually converge on a representative set of states, but the size of this set is large and would scale exponentially in system size. There is a possible fix here, as we expect this system to have a finite correlation length at finite temperature. In theory, if this finite correlation length exists, then one should be able to bound the number of representative states in the MEGA by the size of the Hilbert space on a region proportional to the given correlation length. It would still be an open question as to whether an efficient state preparation scheme is feasible for the situations where the system exhibits a finite correlation length. We leave further analysis of this situation to future work.

Finally, we examine how well the MEGA would ideally work with more generic nonintegrable systems. This is achieved by working in a regime with $U/t=3, U^{\prime}/t = 1.5$ and $t^{\prime}/t=0.75$, where $U^{\prime}$ and $t^{\prime}$ are the strengths of the integrability breaking terms defined in Eq (21). We examine the behavior of this system again on a 10-site ring, with a filling now of $n=0.3$. 

From Fig.~\ref{fig:	Gen_ETH}, we see that the scatter plots of the expectation values of relevant observables do not pinch down to a smooth single-valued function, as they did in the spin band above. When we are away from the edges of the spectrum, we see that most of the points clump together within a small energy range, and we see a decrease in the density if we move vertically away from this point. The ETH conjectures that these fluctuations scale inversely with the density of states, so as we move to large system sizes, we would expect the cloud to become narrower, approaching a single-valued function of the eigenstate energy. One can also see that there are non-typical states in Fig.~\ref{fig:	Gen_ETH} such as the states that have zero double occupancy in the middle of the spectrum. We do not know if they persist in the thermodynamic limit in the strong vs. weak ETH sense~\cite{kim2014testing}.
 
  \begin{figure}[h!]
 	\centerline{
		\includegraphics[scale=0.4]{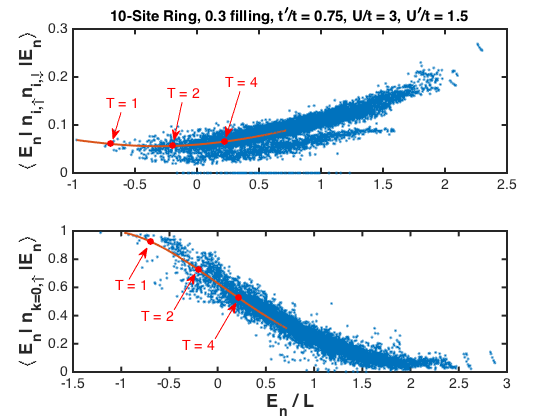}
		}
		\caption{Scatter plots for expecation values of double occupancy and $k=0$ momentum occupation with respect to each energy eigenstate. Here we use a 10-site ring at 0.3 filling, with Hamiltonian parameters of $t^{\prime}/t =0.75, U/t=3.0, U^{\prime}/t=1.5$. As expected by the ETH ansatz, typical eigenstates in the bulk of the spectrum have small fluctuations within a narrow energy range.} \label{fig:		Gen_ETH}
		\end{figure}
 
\begin{figure}[h!]
 		\centerline{
		\includegraphics[scale=0.4]{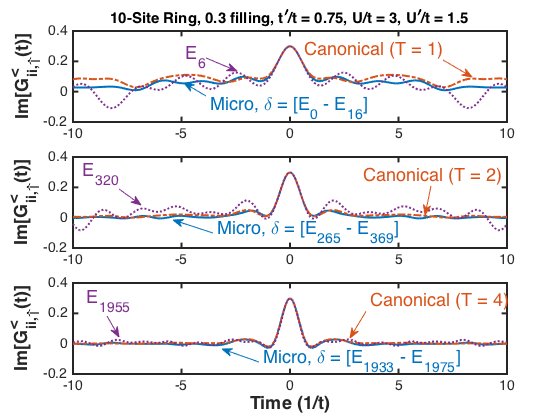}
		}
		\caption{Imaginary part of the local greater Green's function calculated with the microcanonical ramp state (blue curve) and with the exact canonical Gibbs state (red curve) and a single eigenstate (dotted purple curve). As one would expect from a system obeying the ETH, moving to larger temperatures allows the canonical Gibbs states to be approximated by an ensemble of a few or even a single energy eigenstate.} \label{fig:	Gen_Less}
		\end{figure}

For this model, we examine three areas of the energy spectrum corresponding to temperatures of $T=1.0, 2.0,4.0$. For each temperature, Fig.~\ref{fig:	Gen_Less} shows plots of ${\rm Im}[G^<_{ii,\uparrow}(t)]$ calculated with respect to the canonical Gibbs state, a microcanonical window, and a single eigenstate, where each gives the same average energy. At $T=1$ we can see from Fig.~\ref{fig:		Gen_ETH} that the energy eigenstates are sparsely populated in this regime. As a result both the microcanonical and single eigenstate Green functions have trouble converging past $t\approx 1$. As we move to a temperature of $T=2.0$ the spectrum has now become a little bit more dense. As a result the microcanonical ensemble with a large enough window converges rather well, and the single eigenstate holds for a slightly longer period of time before deviating from the canonical Gibbs state result. When we reach a temperature of $T=4$ the spectrum has become rather dense. Here, a smaller microcanonical ensemble converges even better than at $T=2$, even with a smaller number of eigenstates in its energy window, and even a single eigenstate has converged rather well even for late times. The better convergence at large temperature is what is expected from the ETH ansatz, as the entropy and density of states is much larger at higher temperatures. 

These results demonstrate the potential effectiveness of the MEGA in nonintegrable systems. Here, the number of states needed in a MEGA ensemble is inversely proportional to both the size of the system and the temperature. We see that at large enough temperature/system size a MEGA ensemble of even a single eigenstate can reproduce proper thermal properties, as one would expect from the ETH framework. 

There is an ultimate lower bound on the temperature MEGA is able to achieve, as ETH is restricted to eigenstates that have a finite energy density. Most physical systems have either finite energy gaps or algebraically decaying energy gaps as a function of system size. This leads to a zero energy density in the large system limit, effectively allowing large fluctuations in the matrix elements of physical observables of relevant energy eigenstates in this temperature regime . The low-temperature bound here may again potentially be alleviated if this system exhibits a finite correlation length as previously discussed. 
 
\section{conclusions}
  We have outlined the MEGA protocol as a technique to examine the thermal properties of typical observables on quantum computers, and demonstrated its viability using exact diagonalization on small clusters. The advantages of MEGA are in its simplicity to implement and the efficient use of available qubits. While MEGA does not allow one to initially dial in a specific temperature, with an initial guess of a single of set of pure states, one can extract the effective temperature of the system that is represented by those states. Usually one has a rough idea as to within what energy range a typical state lies, such as a Hartree-Fock approximation, for what qualifies as low energy. Normally, one would not know \textit{a priori} whether MEGA can be employed with a small finite set of states. Here, one can simply implement the MEGA protocol and examine how quickly $\beta$ and $\mu$ converge. 
  
 We also showed numerically how systems that obey the ETH are well suited for MEGA, in the appropriate temperature regimes. Beyond this, temperature dependent finite correlation lengths should also theoretically bound the number of states used by MEGA, when they exist, which we hope to examine more thoroughly in the future. In these regimes the size of the MEGA ensemble should ideally be inversely proportional to temperature and system size. The efficiency of MEGA is still limited by system size here as the state preparation procedures and time evolution will scale polynomially with system size. We also recognize that there are certain energy windows where state preparation will scale exponentially, but known Gibbs state preparation/sampling algorithms are inefficient here as well. 
 
Future work may include examining different types of correlators and possibly identifying specific properties in the feedback process to inform what the next ideal state should be to ensure faster convergence.  While the MEGA is not well suited for current quantum hardware, it may be implemented on next generation or "near" term machines once they are capable of handing the circuit depths required for modest time evolution.

~

\acknowledgements

This work was supported by the National Science Foundation under grants numbered PHY-1620555.  JKF was also supported by the McDevitt bequest at Georgetown University. 
\section*{References}
\bibliographystyle{apsrev4-1.bst}
\bibliography{keth}

\end{document}